\documentclass{aa} 
\usepackage{graphicx}
\begin{document}
\def\ffam  {\hbox{$\,.\!\!^{\prime}$}}
\def\ffas  {\hbox{$\,.\!\!^{\prime\prime}$}}
\def\ffM   {\hbox{$\,.\!\!^{\rm M}$}}
\def\ffm   {\hbox{$\,.\!\!^{\rm m}$}}
\def\ffs   {\hbox{$\,.\!\!^{\rm s}$}}

\title{Water vapor in the starburst galaxy NGC253: A new nuclear maser?}

\author{C. Henkel\inst{1} 
        \and
        A. Tarchi\inst{2,3}
        \and
        K.M. Menten\inst{1}
        \and
        A.B. Peck\inst{4}
        }

\offprints{C. Henkel,
\email{chenkel@mpifr-bonn.mpg.de}}

\institute{
           Max-Planck-Institut f{\"u}r Radioastronomie, Auf dem H{\"u}gel 69, 
           D--53121 Bonn, Germany
           \and
           Istituto di Radioastronomia, CNR, Via Gobetti 101, 40129 Bologna, 
           Italy
           \and
           Osservatorio Astronomico di Cagliari, Loc. Poggio dei Pini, Strada
           54, 09012 Capoterra (CA), Italy
           \and
           SAO/SMA Project, PO Box 824, Hilo, HI 96721, USA 
}

\date{Received date / Accepted date}

\abstract{22\,GHz water vapor emission was observed toward the central region 
of the spiral starburst galaxy NGC~253. Monitoring observations with the 
100-m telescope at Effelsberg and measurements with the BnC array of the VLA 
reveal three distinct velocity components, all of them blueshifted with respect
to the systemic velocity. The main component arises from a region close to
the dynamical center and is displaced by $<$1\arcsec\ from the putative nuclear 
continuum source. The bulk of the maser emission is spread over an area not 
larger than 70$\times$50\,mas$^{2}$. Its radial velocity may be explained   
by masing gas that is part of a nuclear accretion disk or of a counterrotating 
kinematical subsystem or by gas that is entrained by the nuclear superwind or 
by an expanding supernova shell. A weaker feature, located $\sim$5\arcsec\ to 
the northeast, is likely related to an optically obscured site of massive star 
formation. Another maser component, situated within the innermost few 10\arcsec\ 
of the galaxy, is also identified.
}

\titlerunning{H$_2$O masers in NGC~253}
\authorrunning{Henkel et al.}

\maketitle

\section{Introduction}

Extragalactic H$_2$O masers, observed in the 6$_{16}$$\rightarrow$$5_{23}$ 
transition at 22.23508\,GHz ($\lambda$$\sim$1.3\,cm), are best known as a 
means to probe accretion disks in Seyfert galaxies (e.g. Miyoshi et al. 
1995; Herrnstein et al. 1999). More than 30 luminous `megamasers' with 
(isotropic) luminosities $L_{\rm H_2O}$ $>$ 10\,L$_{\odot}$ are known to date 
(e.g. Braatz et al. 1996; Greenhill et al. 2003). All of those studied with 
high angular resolution are located within a few parsecs of the nucleus of 
their parent galaxy, tracing either a circumnuclear accretion disk or hinting 
at an interaction between the nuclear radio jet(s) and an encroaching molecular 
cloud (for the latter, see Peck et al. 2003). 

Not all of the known extragalactic water vapor masers show extremely high 
luminosities. Weaker masers, the `kilomasers', were detected in M~33, M~82,
IC~10, NGC~253, M~51, IC~342, NGC~2146, and NGC~6300 (Churchwell et al. 
1977; Huchtmeier et al. 1978; Claussen et al. 1984; Henkel et al. 1986; Ho 
et al. 1987; Becker et al. 1993; Tarchi et al. 2002a,b; Greenhill et al. 2003) 
with (isotropic) luminosities up to $L_{\rm H_2O}$ $\sim$ 10\,L$_{\odot}$.  

Extragalactic H$_2$O masers provide important information about their parent 
galaxies. Studies of `disk-masers' yield mass estimates of the nuclear 
engine and, for NGC\,4258, a calibration of the cosmic distance scale. 
`Jet-masers' provide insights into the interaction of nuclear jets with dense 
warm molecular material and help to determine the speed of the material in the 
jets. Masers in the large scale galactic disks mark locations of ongoing 
massive star formation and can be used to determine, through measurements of 
proper motion, distances to (Greenhill et al. 1993) and three dimensional 
velocity vectors of (Brunthaler et al. 2003) these galaxies. 

Most kilomasers, i.e. those in IC~10, M~33, IC~342, NGC~2146 and M~82, are 
known to be associated with star forming regions (e.g. Argon et al. 1994; 
Baudry \& Brouillet 1996; Tarchi et al. 2002a,b). In M~51, however, the 
maser components arise within 0\ffas 25 from the nucleus, possibly being 
related to the receding jet or to an accretion disk (Hagiwara et al. 2001). 
Although it was detected a long time ago (Ho et al. 1987), the H$_2$O maser 
emission in the prototypical starburst galaxy NGC~253 had not yet been 
studied with high angular resolution. Here we report the first Very Large 
Array (VLA\footnote{The National Radio Astronomy Observatory (NRAO) is 
operated by Associated Universities, Inc., under a cooperative agreement 
with the National Science Foundation.}) observations of the maser(s) in 
NGC~253, complemented by monitoring measurements with the 100-m telescope 
at Effelsberg\footnote{The 100-m telescope at Effelsberg is operated by 
the Max-Planck-Institut f{\"u}r Radioastronomie (MPIfR) on behalf of the 
Max-Planck-Gesellschaft}.

\section{Observations}

\subsection{VLA observations and image processing}

NGC~253 was observed with the VLA in its hybrid CnB configuration on 2002, 
September 22. A frequency setup with a bandwidth of 12.5\,MHz was employed; 
the resulting channel spacing is 2.63\,km\,s$^{-1}$. With 63 channels, a 
total range of 166\,km\,s$^{-1}$ was covered, centered at an LSR velocity 
of 120\,km\,s$^{-1}$. Typically, 3 minute observations of NGC~253 were 
alternated with observations of the phase calibrator 01205--27014. Absolute 
amplitude calibration was obtained using 3C48. The phase center for the NGC~253 
observations was $(\alpha,\delta)_{2000}$ = 00$^{\rm h}$ 47$^{\rm m}$ 
33\ffs 1805, --25$^{\circ}$ 17\arcmin\ 15\ffas 991. The so-called `channel 
0' data, comprising 75\% of the passband, were used to produce a map of the 
continuum plus H$_2$O maser emission. Several iterations of self calibration 
resulted in a high quality map. The final phase and amplitude corrections were 
applied to conventionally calibrated data. The average of the line-free channels 
was then subtracted in the $uv$-plane, resulting in a line-only database, from 
which images of the line emission were made. A continuum image was made from 
the line-free channels. All data reduction was performed using NRAO's 
Astronomical Image Processing system (AIPS). The restoring beam size was 
1\ffas 1 $\times$ 0\ffas8 with a position angle East of North of 88$^{\circ}$.

\subsection{Effelsberg observations}

Observations with the 100-m telescope at Effelsberg were obtained between 
January 1994 and September 2002. The beam size was 40\arcsec. Until 1997, 
observations were carried out in a position switching mode with a K-band maser 
receiver. Later, data were taken with a K-band HEMT receiver in a dual beam 
switching mode with a beam throw of 2\arcmin\ and a switching frequency of
$\sim$1\,Hz. Due to low elevations ($\sim$12$^{\circ}$), the system temperature, 
including atmospheric contributions, was 50--180\,K on an antenna temperature 
scale ($T_{\rm A}^*$; beam efficiency: $\eta_{\rm b}$$\sim$0.3). Flux calibration 
was obtained measuring W3(OH) (3.2\,Jy; Mauersberger et al. 1988). Although gain 
variations of the telescope as a function of elevation were taken into account 
(Eq.\,1 of Gallimore et al. 2001), calibration at low elevations may be uncertain 
by $\pm$30\%. Pointing on nearby continuum sources was found to be better than 
10\arcsec.

\begin{figure}[t]
\vspace{-0.0cm}
\hspace{-0.0cm}
\resizebox{8.8cm}{!}{\rotatebox[origin=br]{0}{\includegraphics{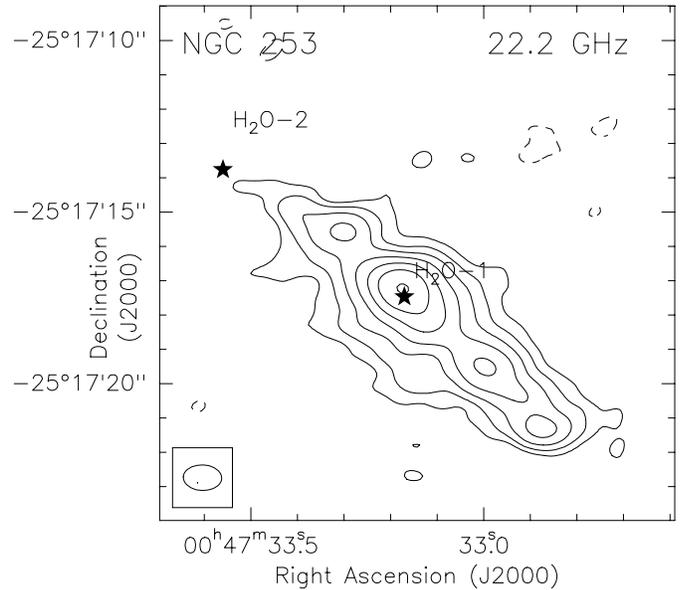}}}
\vspace{0.0cm}
\caption{Map of the 22.2\,GHz continuum emission (epoch: Sept. 22, 2002). Contour 
values are --2, --1, 1, 2, 4$\ldots32$ times the 3$\sigma$ noise level of 
0.9\,mJy\,beam$^{-1}$. The positions of the water masers H$_2$O--1 and H$_2$O--2 
are indicated. Size and orientation of the restoring beam are shown in the 
lower left corner. 
\label{fig1}}
\end{figure}

\section{Results}

Fig.\,1 shows the 22.2\,GHz continuum map which, accounting for the different 
angular resolutions, is consistent with the map of Ulvestad \&\ Antonucci 
(1997, hereafter UA97) at the same frequency. We used the AIPS task SAD to 
search all of the continuum-subtracted channel maps for emission above 
18\,mJy\,beam$^{-1}$, which is 4 times the $1\sigma$ rms noise level. 
H$_2$O emission originates from two locations (see Figs.\,1 and 2): H$_2$O--1 
($V_{\rm LSR}$ $\sim$ 120\,km\,s$^{-1}$) is at $(\alpha, \delta)_{2000}$ = 
$$
 00^{\rm h} 47^{\rm m} 33\,.\!\!^{\rm s}1701 \pm 0\,.\!\!^{\rm s}0004, 
 -25^{\circ} 17\arcmin\ 17\,.\!\!^{\prime \prime}465 \pm 0\,.\!\!^{\prime \prime} 004 
$$
or $(\alpha, \delta)_{1950}$ = 
$$
 00^{\rm h} 45^{\rm m} 05\,.\!\!^{\rm s}7897 \pm 0\,.\!\!^{\rm s}0004, 
 -25^{\circ} 33\arcmin\ 39\,.\!\!^{\prime \prime}474 \pm 0\,.\!\!^{\prime \prime} 004
$$ 
and has an integrated H$_2$O luminosity of 5.62\,Jy\,km\,s$^{-1}$ 
($L_{\rm H_2O}$ $\sim$ 0.8\,L$_{\odot}$). H$_2$O--2 ($V_{\rm LSR}$ 
$\sim$ 170\,km\,s$^{-1}$) is at $(\alpha, \delta)_{2000}$ = 
$$
 00^{\rm h} 47^{\rm m} 33\,.\!\!^{\rm s}559 \pm 0\,.\!\!^{\rm s}004, 
 -25^{\circ} 17\arcmin\ 13\,.\!\!^{\prime \prime}76 \pm 0\,.\!\!^{\prime \prime}03 
$$
and has an integrated H$_2$O luminosity of 145\,mJy\,km\,s$^{-1}$
($L_{\rm H_2O}$$\sim$0.02\,L$_{\odot}$). The emission from H$_2$O--2 is detected 
at a 6--8$\sigma$ level depending on whether the absorption feature next to the 
line (upper panel of Fig.\,2) is included or flagged (bad channel) when determining 
the noise level. The position errors given above are formal values determined using 
the AIPS task JMFIT on a map of integrated emission, in the case of H$_2$O--1 from 
88\,km\,s$^{-1}$ to 149\,km\,s$^{-1}$ and for H$_2$O--2 from 167\,km\,s$^{-1}$ to 
174\,km\,s$^{-1}$. We estimate the {\it absolute} position uncertainty to be 
$\sim$$0\ffas 5$. 

Fig.\,3 presents a close-up of the region around NGC 253 H$_2$O--1, demonstrating
that the bulk of the emission is spread over an area not larger than 70$\times$50 
milliarcseconds (mas), corresponding to 0.8$\times$0.6\,pc ($D$=2.5\,Mpc). Shown 
as squares are relative maser positions from JMFIT with error bars.\footnote{While 
formal errors returned from some software packages have to be viewed with caution, 
we are convinced that JMFIT returns meaningful errors, derived using the rms noise 
around the fitted position. The position error in a coordinate from Gaussian fitting 
is theoretically $\sim$0.5$\theta_{\rm B}/SNR$, where $\theta_{\rm B}$ is the beam 
size and SNR is the map's signal-to-noise ratio (Reid et al. 1980). Using this formula, 
we `manually' calculated fitting errors for a few components and found good agreement 
with the values returned by JMFIT.} The numbers give channel offsets (in units of 
2.63\,km~s$^{-1}$). Position 1 corresponds to 120 km~s$^{-1}$. The fitted flux 
density in all channels is $>$5 times the rms noise of 4.4\,mJy~beam$^{-1}$. Low level 
($<$5$\sigma$) emission with offsets $<$0\ffas2 is also observed but here the positions 
have such error bars that the larger offsets from the reference position may not be 
significant.

\begin{figure}[t]
\centering
\resizebox{7.8cm}{!}{\rotatebox[origin=br]{0}{\includegraphics{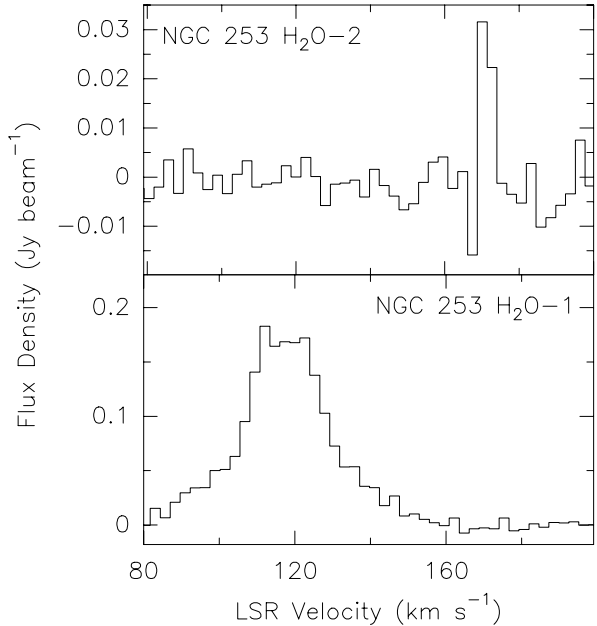}}}
\caption{Lower panel: VLA spectrum, taken on Sept. 22, 2002, of the centrally 
located water maser associated with NGC~253 H$_2$O--1. Upper panel: VLA spectrum 
of the north-eastern water maser associated with NGC~253 H$_2$O--2. 
The velocity scale is LSR (Local Standard of Rest). $V_{\rm HEL}$ 
= $V_{\rm LSR}$ + 7.1\,km\,s$^{-1}$. For both spectra, the channel spacing is 
2.63\,km\,s$^{-1}$. 
\label{fig2}}
\end{figure}

\begin{figure}[t]
\vspace{-2.5cm}
\hspace{-0.8cm}
\resizebox{11.2cm}{!}{\rotatebox[origin=br]{0}{\includegraphics{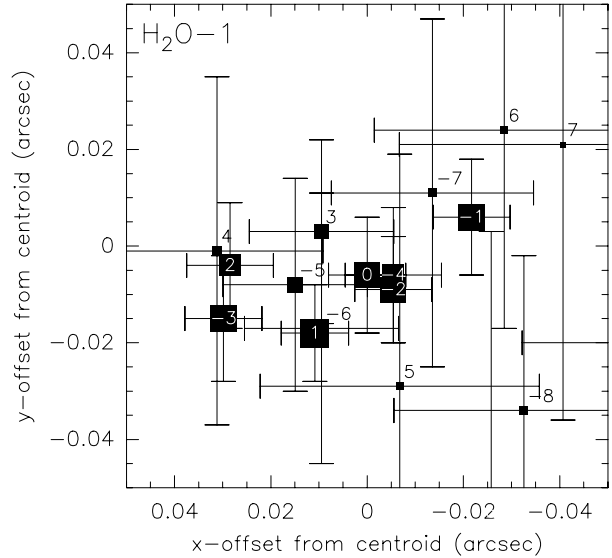}}}
\vspace{-6.5cm}
\caption{Close-up of the region around NGC 253-H$_2$O--1. Shown as squares 
are the maser positions from JMFIT with error bars. Position offsets in 
($\alpha_{2000},\delta_{2000}$) are relative to the centroid position given 
in Sect.\,3. The numbers give velocity offsets in units of 2.63\,km~s$^{-1}$ 
from 117.37\,km~s$^{-1}$, i.e. position 1 corresponds to 120\,km\,s$^{-1}$. 
The side lengths of the symbols are proportional to the emission in that channel 
(187\,mJy~beam$^{-1}$ for position 1).
\label{fig3}}
\end{figure}

\begin{figure}[t]
\vspace{-1.0cm}
\resizebox{20.0cm}{!}{\rotatebox[origin=br]{-90}{\includegraphics{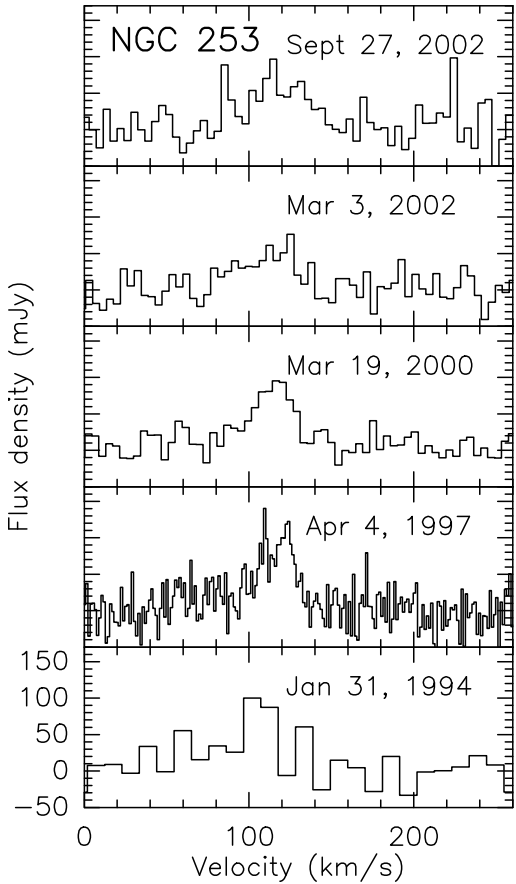}}}
\vspace{-1.7cm}
\caption{H$_2$O spectra with LSR velocity scale taken between 1994 and 2002.
No redshifted emission was seen beyond 260\,km\,s$^{-1}$ out to 400\,km\,s$^{-1}$
(Jan. 1994, Apr. 1997), to 950\,km\,s$^{-1}$ (Mar. 2000), to 800\,km\,s$^{-1}$ 
(Mar. 2002) and to 400\,km\,s$^{-1}$ (Sept. 2002). Channel spacings and 1$\sigma$
noise levels are (upper to lower panel) 4.2, 4.2, 4.2, 1.3, and 10.5\,km\,s$^{-1}$
and 21.7, 19.2, 11.6, 22.2 and 20.6\,mJy, respectively. The systemic velocity is
$V_{\rm sys}$ $\sim$ 240\,km\,s$^{-1}$ (see Sect.\,3).
\label{fig4}}
\end{figure}

Fig.\,4 shows the Effelsberg spectra covering a time interval of more than 
eight years. Emission from H$_2$O--1, sometimes split into two velocity components, 
dominates all spectra, at a level of $\sim$100\,mJy. This is consistent with 
data collected before (Ho et al. 1987; Nakai \& Kasuga 1988). The agreement 
between the VLA spectrum (Fig.\,2) and that observed with Effelsberg five days 
later (Fig.\,4, upper panel) is reasonably good. While the VLA spectrum is more 
sensitive, the Effelsberg spectrum covers a wider velocity range. The Effelsberg 
spectrum contains a weak tentative feature at 170\,km\,s$^{-1}$ that is convincingly 
detected by the VLA (H$_2$O--2). Differences between the spectra are, however, also 
apparent: In the VLA spectrum the flux density of H$_2$O--1 is twice that 
in the Effelsberg spectrum, while the tentative 85\,km\,s$^{-1}$ component, 
apparent in Fig.\,4, is not confirmed. The Effelsberg spectrum also shows a 
225\,km\,s$^{-1}$ feature (H$_2$O--3) that is close to the systemic velocity 
of the galaxy ($V_{\rm sys}$ $\sim$ 225--245\,km\,s$^{-1}$; e.g. Canzian et al.
1988; Anantharamaiah \& Goss 1996; Prada et al. 1996). Since this feature was not 
known prior to the interferometric mesurements, its velocity range is not part of 
the VLA observations.

\section{Discussion}

The Sculptor galaxy NGC~253 is a highly inclined ($i$$\sim$78$^{\circ}$) nearby 
($D$$\sim$2.5\,Mpc) barred Sc spiral and is one of the brightest sources of 
far infrared emission beyond the Magellanic Clouds (e.g. Pence 1981; Scoville 
et al. 1985; IRAS 1989; Mauersberger et al. 1996; Das et al. 2001). Its 
spectral energy distribution has been studied from radio waves (e.g. UA97) 
to TeV $\gamma$-rays (Itoh 2002). At many wavelengths the nuclear environment 
is dominated by a starburst that is confined to a $\sim$100\,pc region located 
in the plane of the galaxy and that is centered slightly southwest of the dynamical 
center (e.g. Keto et al. 1993, 1999; Telesco et al. 1993; B{\"o}ker et al. 1998). 
Radio continuum measurements indicate that NGC~253 contains a large number of potential 
supernova remnants and H{\sc ii} regions near its center. The strongest of these 
radio sources, located right at the dynamical center, remains unresolved in VLA 
images, shows a relatively flat radio spectral index and has a brightness temperature 
in excess of 10$^{4}$\,K (Turner \& Ho 1985; Antonucci \& Ulvestad 1988; Ulvestad \& 
Antonucci 1991, 1994; UA97). The source is likely to indicate the position of the 
active galactic nucleus (AGN) in NGC~253.

\subsection{H$_2$O--1}

The position of H$_2$O--1 can be compared with the positions of nine radio 
continuum sources that were detected in the VLA A configuration by UA97 at 
$\lambda$=1.3\,cm. Three of the sources are close to the maser, TH2, TH3 and 
TH4. TH2 is nominally offset from H$_2$O--1 by ($\Delta \alpha, \Delta 
\delta$) = ($+0\ffas 06, +0\ffas 38$), TH3 by ($+0\ffas 00, -0\ffas 29$) and 
TH4 by ($-0\ffas 10, +0\ffas 04$). TH2 is the strongest of the 64 radio sources 
compiled by UA97. It is the source that is associated with the putative AGN. 
TH4 may be a supernova remnant, while the spectrum of TH3 is not well determined. 

Our absolute position uncertainty is 0\ffas 5 (Sect.\,3). The registration 
of our maps with those of UA97 obtained with different phase calibrators and the 
conversion of B1950.0 to J2000.0 coordinates (using NED\footnote{This research 
has made use of the NASA/IPAC Extragalactic Database (NED) which is operated by 
the Jet Propulsion Laboratory, California Institute of Technology, under contract 
with the National Aeronautics and Space Administration.}) may further increase 
the total error budget, but only slightly. Comparing the coordinates given by 
Turner \& Ho (1985) and UA97 for the hypothesized nuclear continuum source, we 
find a difference of 0\ffas 15, while the coordinate conversion should introduce 
an error of order 0\ffas 05 (e.g. Fricke 1982; Ma et al. 1998). 

H$_2$O--1 appears to be closest to TH4, but TH2 and TH3 are also located within 
the region defined by our absolute position uncertainty. H$_2$O--1 is then not 
necessarily associated with TH4. An association with TH2 or TH3 is also possible. 
We thus conclude that {\it the maser arises from the nuclear region of the galaxy. 
A direct association with the nucleus is not ruled out by our measurements}.

Having identified a maser component close to the hypothesized galactic nucleus,
its lineshape, radial velocity and association with other sources deserve to be 
briefly discussed. H$_2$O--1 shows a smooth broad profile that is rarely seen 
in galactic or extragalactic H$_2$O maser sources. The profile resembles those 
seen toward the jets of AGN (i.e. NGC~1052 and Mrk~348; Braatz et al. 1996; 
Peck et al. 2003) and toward prominent star forming regions in M~33 and 
NGC~2146 (Huchtmeier et al. 1978; Tarchi et al. 2002b). The profile does {\it 
not} resemble any lineshape observed toward circumnuclear disks (e.g. NGC~4258,
Mrk~1419 and NGC~1068; Miyoshi et al. 1995; Gallimore et al. 2001; Henkel et al. 
2002). In view of the small number of sources studied thus far, however, the 
strength of this argument is difficult to assess. 

H$_2$O--1 appears to be located slightly southwest of the nucleus, i.e. on 
that side of the edge-on galaxy that is redshifted (e.g. Fig.\,1 of Das et al. 
2001). If we adopt this view, the blueshift of $\sim$120\,km\,s$^{-1}$ w.r.t. 
$V_{\rm sys}$ appears to be highly peculiar. The enormous discrepancy between 
observed and expected velocity that is of order 200--300\,km\,s$^{-1}$ can 
then be explained by the following possibilities: (1) The maser is part of a 
circumnuclear acretion disk, likely centered at TH2; (2) the emission originates
from a counter-rotating nuclear core component; (3) the masing gas is entrained
by an expanding supernova shell or (4) the masing gas is entrained by the nuclear 
wind. 

If the maser is tracing an accretion disk around a supermassive central object 
as in NGC~4258, we may exclusively view the blueshifted side of this disk that may 
be displaced by a few milliarcseconds from the dynamical center. Such a position is 
well within the limits of our observational accuracy. The presence of a prominent 
source of hard X-rays at the very center and nuclear column densities of several 
10$^{23}$\,cm$^{-2}$ (Weaver et al. 2002) are consistent with such a scenario.

From observations of the 8.3\,GHz H\,92$\alpha$ radio recombination line with a 
resolution of 1\ffas 8$\times$1\ffas 0, Anantharamaiah \& Goss (1996) identify three 
kinematic subsystems near the center of NGC~253, one with solid body rotation in the 
same sense as the galactic large scale disk, a second exhibiting rotation in a plane 
perpendicular to this disk and a third one in the innermost 2\arcsec\ that may 
actually counterrotate. H$_2$O--1 is located within this innermost region.

An association with the envelope of a rapidly expanding supernova remnant (e.g.
TH4) interacting with a dense molecular cloud might explain the peculiar velocity 
of H$_2$O--1. Such sources are, however, not common in the Galaxy, likely because gas 
densities are too low (Claussen et al. 1999). M~82, a nearby starburst galaxy like
NGC~253, also contains a large number of supernova remnants (e.g. Kronberg et al. 1985). 
The H$_2$O masers appear, however, to be associated with prominent star forming regions 
(Baudry \& Brouillet 1996). While the parent molecular clouds are often located at 
the periphery of supernova remnants, a direct association with their expanding shells 
is not obvious. While conditions near the nucleus of NGC~253 may differ from those
in the Milky Way and M~82, a scenario involving expanding supernova shells is therefore 
not likely.

A nuclear plume is seen in the far infrared and X-ray continuum (e.g. Rice 1993;
Pietsch et al. 2000), in optical emission lines (e.g. Heckman et al. 1990; Schulz 
\& Wegner 1992), and in the $\lambda$18\,cm lines of OH (Turner 1985). Most of the 
gas in the OH plume, that is most prominent north of the nucleus, is receding with 
respect to $V_{\rm sys}$ and is likely arising from the far side of the galaxy disk. 
Therefore the base of this OH plume should not be blueshifted w.r.t. $V_{\rm sys}$, 
which would be necessary in order to be compatible with the blueshifted H$_2$O 
component observed towards H$_2$O--1. A direct association between H$_2$O--1 and 
the OH plume is thus unlikely.

Radio jets can be a strong agent to drive maser action when hitting high density
clumps or entraining molecular gas (e.g. Elitzur 1995; Peck et al. 2003). Less
violent, optically detectable ionized outflowing material might, however, also 
trigger maser emission (Schulz \& Henkel 2003). In NGC~253, optical emission lines 
and the X-ray continuum indicate the presence of a cone-like structure that is 
viewed almost edge-on and that originates from the nuclear region (Pietsch et al. 
2000 and references therein). Likely representing a superwind triggered by the 
starburst, the cone surface may indicate the boundary between the outflowing gas 
and the ambient medium. The cone is seen southeast of the nuclear region, i.e. on 
the opposite side of the OH plume, and shows blueshifted emission. This cone is, 
however, observed at distances of several arcseconds SE of the nucleus and its 
presence and nature in the very nuclear region remains an enigma. We conclude 
that it is not obvious how an interaction of this wind with a molecular cloud 
could explain the line emission from H$_2$O--1.

From analogy to the Galaxy, an association of H$_2$O--1 with a red giant star 
can be excluded on the basis of the high luminosity of the maser. A star forming 
region with a broad dominant maser component that does not strongly vary with 
time and that shows no narrow occasionally flaring spikes is also not known in 
the Galaxy. 

{\it To summarize} the discussion on H$_2$O--1, we find that an association with the 
nuclear source is an attractive scenario, while a connection with a counterrotating 
kinematical subcomponent, an expanding supernova remnant or the superwind cannot be 
ruled out. Among possible counterparts to H$_2$O--1, we should also consider compact 
regions of molecular line emission. At first sight, an association between the 18\,cm 
OH satellite lines masers (region 7 according to Frayer et al. 1998) and H$_2$O--1 may 
appear farfetched since the main OH component has a radial velocity of 
$V$$\sim$210\,km\,s$^{-1}$. Weaker OH emission is, however, seen between 100 and 
200\,km\,s$^{-1}$ so that an association {\it might} be possible. 

\subsection{H$_2$O--2}

H$_2$O--2 is situated northeast of the nucleus. Here the radial velocity is 
consistent with its position (see e.g. Paglione et al. 1995). An association 
with young massive stars is likely. The maser may arise from an optically 
highly obscured region. No X-ray, optical, near infrared or radio counterpart is 
apparent (e.g. Kalas \& Wynn-Williams 1994; UA97; Dudley \& Wynn-Williams 1999;
Vogler \& Pietsch 1999). The maser may be located in a region of intense SiO 
emission (see the SiO 171.25\,km\,s$^{-1}$ panel in Fig.\,3 of Garc\'{\i}a-Burillo 
et al. 2000). SiO is a tracer of shock chemistry that can also enhance H$_2$O 
abundances by several orders of magnitude.

\section{Concluding remarks}

Obviously, A configuration VLA data are needed to determine more accurate relative 
positions between the dominant 22\,GHz H$_2$O maser (H$_2$O--1) and the dominant 
radio continuum source of the galaxy. This would allow us to confirm or to reject
a direct association between H$_2$O--1 and the putative AGN. Such measurements 
should also include an attempt to spatially resolve the emission from H$_2$O--1 
and a determination of the position of the near systemic feature detected in the 
most recent Effelsberg spectrum. In any case, the small projected distance between 
H$_2$O--1 and the central radio source shows that the maser is arising from the 
nuclear region of NGC~253. As early as 1987, Ho et al. postulated the existence of a 
numerous family of weak nuclear masers. This family was supposed to form the low 
luminosity tail of a megamaser distribution that is likely characterized by highly 
non-isotropic emission. Such masers are difficult to detect because they are too weak 
to be observed at distances much in excess of 10\,Mpc. It is possible that after 
identifying weak nuclear maser emission in M~51 (Hagiwara et al. 2001), we have 
found a second such case. Being located at a distance of only $D$$\sim$2.5\,Mpc, 
follow-up studies with high spatial resolution will be possible.

\begin{acknowledgements}
We wish to thank an anonymous referee and J.S. Ulvestad for critical comments and 
are grateful to the VLA and 100-m Effelsberg staff for their cheerful assistance.
\end{acknowledgements}

\end{document}